 \documentclass{aa}  
\usepackage{graphicx}
\usepackage{txfonts}
\usepackage{amssymb}
\begin{document} 

  \title{Mid-infrared interferometry with $K$ band fringe-tracking\thanks{Based on data products from observations with ESO Telescopes at the La Silla 
Paranal Observatory under program ID 087.C-0824, 090.B-0938, and 60.A-9801(M).}}
%    \title{PRIMA FSU-A as a fringe tracker for MIDI}
%    \title{A $K$ band fringe tracker for the MID-infrared Interferometric instrument}
   \subtitle{I. The VLTI MIDI+FSU experiment}

   \author{A.~M\"{u}ller\inst{1,2}
	  \and J.-U.~Pott\inst{2}
	  \and A. M\'{e}rand\inst{1}
          \and R.~Abuter\inst{3}
          \and F.~Delplancke-Str\"{o}bele\inst{3}
          \and Th.~Henning\inst{2}
          \and R.~K\"{o}hler\inst{2}
          \and Ch.~Leinert\inst{2}
          \and S.~Morel\inst{1}
	  \and T.~Phan~Duc\inst{3}
	  \and E.~Pozna\inst{3}
	  \and A.~Ramirez\inst{1}
	  \and J.~Sahlmann\inst{4}
          \and C.~Schmid\inst{3}
          %or henning, delplancke nach autoren fragen
          }

   \institute{European Southern Observatory, Alonso de Cordova 3107, Vitacura, Santiago, Chile\\\email{amueller@eso.org}\and
	      Max-Planck-Institut f\"ur Astronomie, K\"onigstuhl 17, 69117 Heidelberg, Germany\and
	      European Southern Observatory, Karl-Schwarzschild-Str. 2, 85748 Garching b. M\"{u}nchen, Germany \and
European Space Agency, European Space Astronomy Centre, P.O. Box 78, Villanueva de la Ca\~{n}ada, 28691 Madrid, Spain\\%\and
             }

   \date{Received ; accepted }

% \abstract{}{}{}{}{} 
% 5 {} token are mandatory
 
  \abstract
   {A turbulent atmosphere causes atmospheric piston variations leading to rapid changes in the optical path difference of an interferometer, which 
causes correlated flux losses. This leads to decreased sensitivity and accuracy in the correlated flux measurement.}
   {To stabilize the $N$ band interferometric signal in MIDI (MID-infrared Interferometric instrument), we use an external fringe tracker 
working in $K$ band, the so-called FSU-A (fringe sensor unit) of the PRIMA (Phase-Referenced Imaging and Micro-arcsecond Astrometry) facility at VLTI. We present measurements obtained using the newly commissioned and publicly offered MIDI+FSU-A mode. A first characterization of the fringe-tracking performance and resulting gains in the $N$ band are presented. In addition, we demonstrate the possibility of using the FSU-A to measure visibilities in the $K$ band.}
   {We analyzed FSU-A fringe track data of 43 individual observations covering different baselines and object $K$ band magnitudes with respect to the 
fringe-tracking performance. The $N$ band group delay and phase delay values could be predicted by computing the relative change in the differential 
water vapor column density from FSU-A data. Visibility measurements in the $K$ band were carried out using a scanning mode of the FSU-A. 
%    To test the reliability of the FSU-A visibility measurements two binary stars with previously known astrometric orbit were observed. A binary model was fitted to the data to retrieve the astrmetric coordinates.
   }
   {Using the FSU-A $K$ band group delay and phase delay measurements, we were able to predict the corresponding $N$ band values with high accuracy 
with residuals of less than 1\,$\mu$m. This allows the coherent integration of the MIDI fringes of faint or resolved $N$ band targets, respectively. 
With that method we could decrease the detection limit of correlated fluxes of MIDI down to 0.5\,Jy (vs. 5\,Jy without FSU-A) and 0.05\,Jy (vs. 
0.2\,Jy without FSU-A) using the ATs and UTs, respectively. The $K$ band visibilities could be measured with a precision down to $\approx$2\%.}
   {}
  \keywords{Instrumentation: interferometers -- Techniques: interferometric -- Methods: observational -- Methods: data analysis}

   \maketitle
%
%________________________________________________________________

\section{Introduction}
% Every high-resolution technique at optical-infrared wavelengths suffers from atmospheric turbulence, causing piston and tip tilt variations, which degrade the signal quality. To overcome this issue one has to correct in millisecond regime depending on the wavelength regime (coherence time/length). 
% % Additional disturbances such as telescope vibrations caused by instruments with Closed-Cycle Cooler Heads systems or wind exposure can alter the signal to be recorded \citep[e.g.][]{pou10}, too. 
% Fluctuations in the optical path difference (OPD) lead to a degradation of the fringe power, and hence a jitter in visibility can be seen.\\
% To overcome this situation the usage of an external fringe tracker is required to send offsets to a delay line in real-time. Ideally, this will lead to a stable (with respect to variations of the fringe position) interferogram on the beam combiner used for the scientific observation. Besides an increase in data quality (stability of correlated flux measurements) an increase in sensitivity without increasing the integration time can be observed.\\
%%%%%%%%%%%%
Every high-resolution technique at optical-infrared wavelengths suffers from atmospheric turbulence, causing piston and tip tilt variations in the stellar light beams. 
%Additionally, disturbances such as telescope vibrations caused by instruments with Closed-Cycle Cooler Heads systems or wind exposure alter the signal to be recorded (e.g. Poupar et al. 2010). 
For interferometric observations, any fluctuation, particularly in the optical path difference (OPD), leads to a degradation of the fringe power, hence a jitter in the fringe visibility. To overcome these problems, the various disturbances have to be corrected on a time scale of milliseconds -- depending on the observed wavelength.\\
For interferometry such a correction usually implies using an external fringe tracker that sends offsets to a delay line in real time. Ideally, this will lead to a stable (with respect to variations in the fringe position) interferogram on the beam combiner used for the scientific observation. Besides an increase in data quality (stability of correlated flux measurements), an increase in sensitivity without extended integration time can be observed. At the ESO  Very Large Telescope Interferometer (VLTI, \citet{hag12}), we set this up for the MID-infrared Interferometric instrument (MIDI \citet{lei03}) where the PRIMA FSU-A is used as an external fringe tracker, the so-called MIDI+FSU-A mode \citep{mue10,pot12}.\\

MIDI is a two-beam Michelson type interferometer, which operates in the $N$ band, and produces spectrally dispersed fringe packages between 8\,$\mu$m and 13\,$\mu$m. The instrument is usually operated with a prism as dispersive element, providing a resolution of $R\approx30$. Given the wavelength range, MIDI is suited to observations of objects with dusty environments, especially young stellar objects (YSO, e.g., \citet{boe04,rat09,olo13,bol13}) surrounded by circumstellar envelopes and protoplanetary disks and active galactic nuclei (AGN, e.g., \citet{jaf04,bur13}). Since 2004 the MIDI instrument has been available to the astronomical community and become one of the most prolific interferometric instruments to date with more than 100 scientific publications in refereed journals.\\

The new facility at VLTI for phase-referenced imaging and micro-arcsecond astrometry (PRIMA, \citet{del08,bel08}) is currently undergoing its commissioning phase. PRIMA consists of four subsystems: the star-separator module (STS, \citet{nij08}), the internal laser metrology \citep{sch07}, the differential delay lines (DDL, \citet{pep08}), and two identical fringe sensor units (called FSU-A and FSU-B). However, for the MIDI+FSU-A mode alone, the FSU-A from the PRIMA subsystems is relevant. A detailed description of the FSUs is given by \citet{sah09}. We briefly summarize the principle setup of the FSU-A:
\begin{itemize}
 \item The FSU-A operates in the $K$ band between 2.0\,$\mu$m and 2.5\,$\mu$m.
%  \item The beams of the two telescopes enter the FSU-A and going through an alignment and compensation unit (ACU), which consists of flat mirrors mounted on a piezo tip-tilt and translation platform.
%  \item For one beam an achromatic phase shifter introduces a phase shift of 90\degr between $p$ and $s$ polarization.
%  \item The beam combiner superimposes both beams and splits them by 50/50.
 \item Polarizing beam splitters separate the $p$ and $s$ components of the individual beams, resulting in four beams separated by 90\degr in phase, before they are injected into single-mode fibers.
%  \item The fibers send the light inside a cryostat where it gets spectrally dispersed over five pixels per quadrant by a prism and is finally recorded by a detector.
 \item The fibers send the light inside a cryostat where each of the four beams gets dispersed by a prism over five pixels on one quadrant of a detector.
 \item For each quadrant a synthetic white light pixel is computed by summing up the flux measured by the five spectral pixels.
\end{itemize}
Given the setup, the four light beams are phase-shifted spatially. The intensity of a sinusoidal signal (or fringe package) is therefore measured at four points separated by a quarter of a wavelength. This so-called ABCD principle allows computation of the fringe phase and group delay, as well as visibility directly (see Eqs.~4 and 5 in \citet{sha77}) and in real time.\\

In this paper, the operation scheme of the MIDI+FSU-A mode is briefly described in Sec.~\ref{sec:operation}. The performance of this new mode and the measurement of $K$ band visibilities are presented in Sects.~\ref{sec:performance} and~\ref{sec:vis}, respectively. Discussion and conclusions can be found in Sec.~\ref{sec:conc}.\\

In the next paper of this series, we will discuss the achievable differential phase precision in the MIDI+FSU mode in detail, which is significantly improved by factors of 10--100 over the MIDI stand-alone operation thanks to the synchronous FSU tracking information. This new capability will be discussed in the context of faint companion detections down to the planetary regime.

\section{Operating MIDI+FSU-A}\label{sec:operation}

The MIDI+FSU-A mode was made available to the community in April 2013 (ESO period 91), already. Therefore, it is part of a standardized observing scheme using observing templates, allowing efficient observations. Night operations at VLT/I consist of executions of observing blocks (OB) that are made of templates. Each template is tuned to a particular observation by setting its parameters called ``keywords'' \citep{per97}. The degree of complexity of the new acquisition template compared to the MIDI acquisition template increased only slightly with respect to available instrument parameters. The new keywords include flags for FSU-A sky calibration, operation frequency, and setups for fringe search and fringe scanning. The observing template has no new keywords.

\subsection{VLTI setup}

The MIDI+FSU-A mode can be operated with two 8.2\,m unit telescopes (UT) or two 1.8\,m auxiliary telescopes (AT). The light beams of the two telescopes are sent to a tunnel with delay lines (DL) to equalize the light paths of the two telescopes. After the beams enter the VLTI laboratory, a dichroic plate transmits the $N$ band to MIDI while the near-infrared part of the light is transmitted to the other subsystems. Another dichroic plate sends the $H$-band to the IRIS (Infra-Red Image Sensor) tip-tilt corrector \citep{git04}, while the remaining $K$ band is transmitted into the FSU-A. During observations, OPD corrections are sent through the optical path difference controller (OPDC) of the VLTI, which sends offsets to the VLTI delay lines. MIDI therefore only acts as cold camera. Fringe tracking by FSU-A and observations by MIDI occur on the same target, i.e. on-axis.

\subsection{Observation sequence}

In this section we describe a typical observation from a user's point of view with its individual steps.
\begin{itemize}
 \item Presetting (slewing) of the telescopes and delay lines, start of telescope, and IRIS ``lab guiding'' (VLTI tunnel and laboratory tip/tilt 
control); duration $\approx$3\,minutes.
 \item Optional (mostly on the first target of the night that uses a given baseline): check of telescope pupil alignment and beam alignment in MIDI; 
duration $\approx$10\,minutes.
 \item Optimization of flux injection into FSU-A using a beam tracking algorithm; duration $\approx$3\,minutes.
 \item Record of sky background and flat fields for FSU-A; duration $\approx$3\,minutes.
 \item Optional: Fringe search by driving a ramp with the main DL of typically $\pm$1\,cm to measure the zero OPD offset; duration $\approx$1\,minute.
 \item Optional: Fringe scanning through the fringe package for typically 250 times; duration $\approx$4\,minutes.
 \item Initialization of fringe-tracking and MIDI+FSU-A data recording; duration typically $\approx$3\,minutes.
\end{itemize}
If the optical paths are well aligned at the beginning of the night, a pure calibrator--science--calibrator observing sequence can be carried out 
typically in 30 minutes, making the MIDI+FSU-A mode into an efficient instrument even with its increased complexity. For detailed explanations 
about the PRIMA FSU fringe-tracking concept, we refer the interested reader to \citet{sah10} and \citet{sch10}.

% fsu dits
% v,n,k,h needs to be known

%______________________________________________________________

\section{MIDI+FSU-A performance}\label{sec:performance}

The $N$ band covers a range from 8\,$\mu$m to 13\,$\mu$m. This corresponds to a blackbody radiating with its highest flux density at roughly room 
temperature ($\approx$300\,K). Therefore, MIDI is dominated by a large thermal background stemming from sky emission, as well as of approximately 
two dozen mirrors in the light path between the source and the beam combiner. Because the scientific sources on sky are fainter by at least three 
orders of magnitudes with respect to the thermal background, one cannot simply use longer integration times as in the optical spectral regime. 
Otherwise, the detector would saturate. There is therefore a natural sensitivity limit for sources that can be detected by MIDI alone, which does not 
depend on the detector characteristics (except for its full well capacity and the size of the telescope's primary mirror). In addition, atmospheric 
piston turbulence decreases the ability to reliably extract the interferometric signal in post-processing. Using an external fringe tracker, 
such as the FSU-A, provides additional information about the presence (in case the fringe-tracking loop is locked) and position (extrapolation form 
$K$ band) of a fringe package, which can be used for the later reduction process. In the following, we both present several characteristics of the 
MIDI+FSU-A mode and highlight unmatched achievements. 
\\

\begin{figure}[!ht]
  \centering
  \includegraphics[width=9cm]{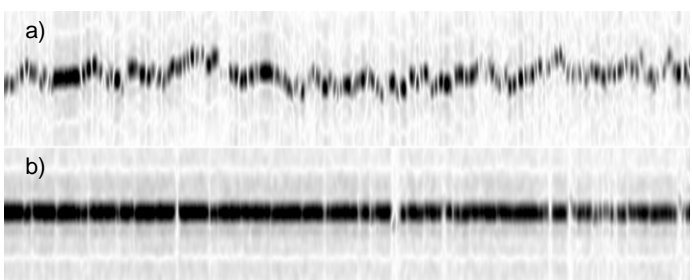}
  \caption{Waterfall plot of MIDI observations without (upper plot) and with FSU-A (lower plot) as an external fringe tracker. The horizontal axis 
can 
be interpreted as time, the vertical axis as OPD. Both plots display 2.5\,min of observations.}
  \label{fig:wfplot}
\end{figure}

In Figure~\ref{fig:wfplot} we show a section of a so-called ``waterfall'' plot\footnote[1]{A ``waterfall'' plot shows the amplitude and position of 
the Fourier transform of a scan through the interferogram with time.} of two typical MIDI observations taken without (Fig.~\ref{fig:wfplot}a) and 
with (Fig.~\ref{fig:wfplot}b) the FSU-A as external fringe tracker. The observations were performed on the same star (HD~187642, Altair) 
during the first MIDI+FSU-A test in the night of July 23, 2009, and were separated by 90 minutes (i.e., observed under almost the same conditions). 
In Fig.~\ref{fig:wfplot}a) the variations introduced by atmospheric piston fluctuations during the course of the MIDI observation can be clearly 
identified. The residuals in group delay (GD) are on the order of 11\,$\mu$m. Using the external fringe tracker the $N$ band signal is stabilized 
well (Fig.~\ref{fig:wfplot}b) and atmospheric piston variations are corrected well by FSU-A. \\

\begin{figure}[!ht]
  \centering
  \includegraphics[width=9cm]{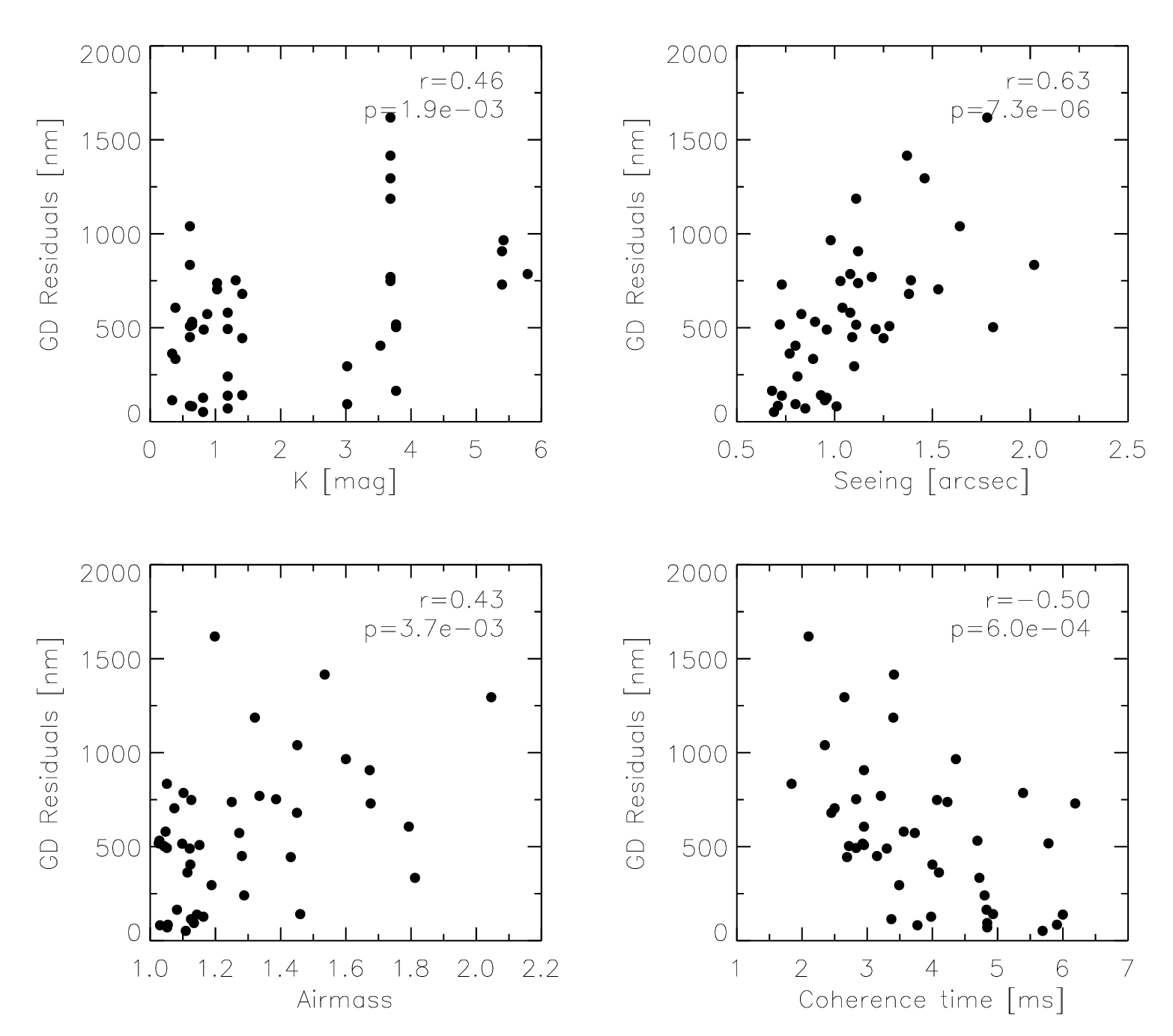}
  \caption{FSU-A group delay residuals as a function of $K$ band magnitude (upper left), seeing (upper right), airmass (lower left), and coherence 
time (lower right).}
  \label{fig:gdres}
\end{figure}

\begin{figure}[!ht]
  \centering
  \includegraphics[width=9cm]{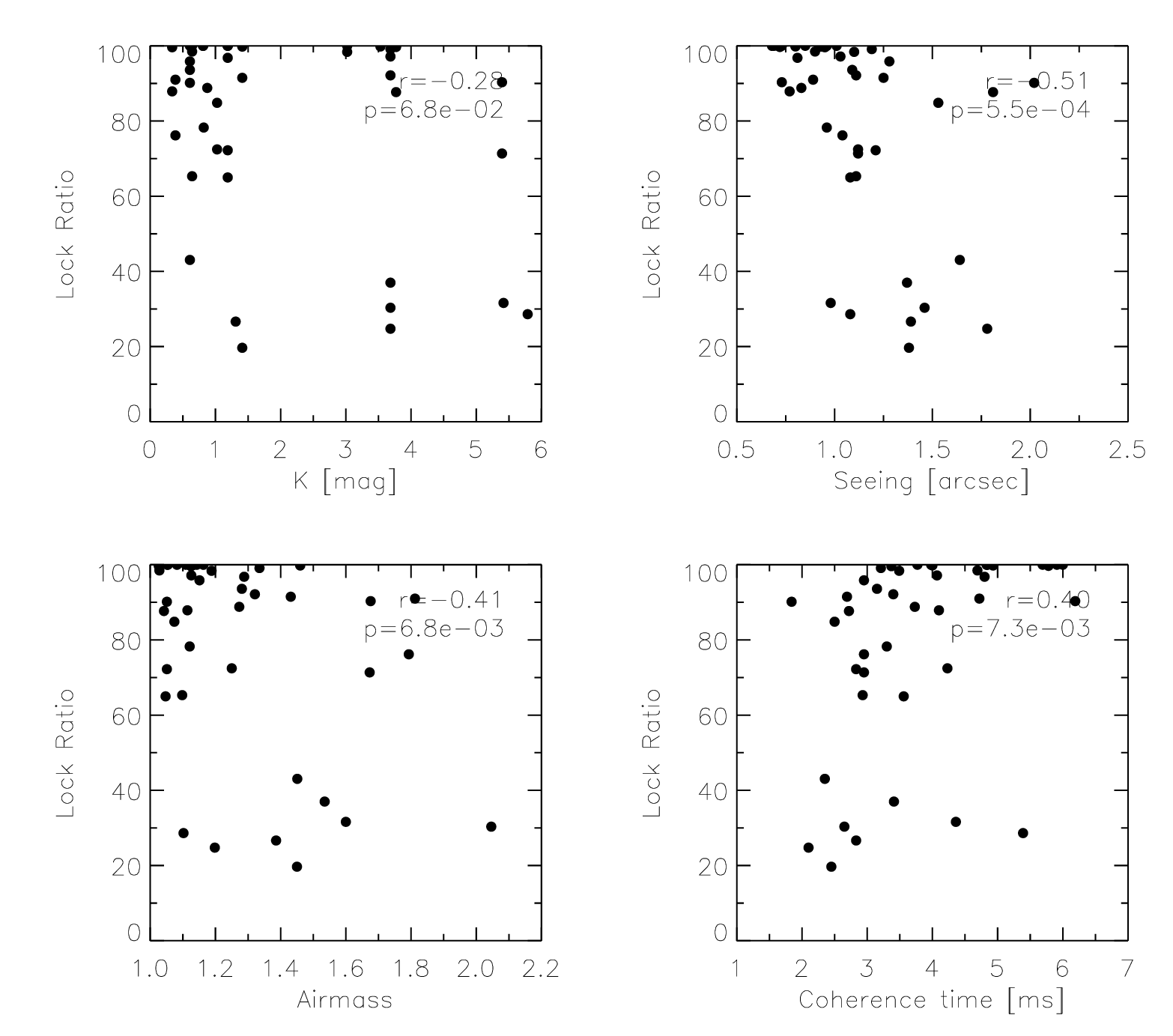}
  \caption{Same as Fig.~\ref{fig:gdres} but for the FSU-A lock ratio.}
  \label{fig:lr7}
\end{figure}

A requirement for an external fringe tracker is the ability to correct for OPD variations caused by a turbulent atmosphere in real time; i.e., 
corrections have to be sent to the delay lines that are not slower than the current coherence time $\tau_0$ of the atmosphere. The average value for 
the coherence time at Paranal is $\tau_0=3.9$\,ms in the visible spectral range at 500\,nm \citep[e.g.,][]{sar02}. Therefore, an external fringe 
tracker has to operate in the kilo-Hertz regime with low latency on the order 1\,ms. In the nights of November 29 to 31 in 2011 we observed various 
objects and 
calibrator stars (43 observations in total, see Table~\ref{tbl:obsmf} for the observing log) during the course of the guaranteed time observations 
under program ID 087.C-0824. The observations were carried out under various conditions and allow a first assessment of the fringe-tracking 
performance. Figure~\ref{fig:gdres} shows the measured FSU-A group delay residuals as a function of $K$ band magnitude, seeing, airmass, and 
coherence time. For each subplot the linear correlation coefficient $r$ and the probability $p$, which indicates whether the measured correlation can 
be generated by a random distribution \citep{bev03}. The FSU-A performs well for the range of $K$ band magnitudes explored here based on our 
requirements of what we consider a useful data set. 
% The FSU-A performs well \textbf{for the here explored} of the $K$ band brightness of the 
% source \textbf{based on our requirements of what we consider a useful data set. 
This has a lock ratio of at least 30\% and group delay 
residuals of less than $\lambda_{N-band}$/10, which is fulfilled for the majority of the data sets. A clear positive correlation can be identified 
for seeing, and a negative correlation for the coherence time. This can be explained by differential flux injection into the ABCD fibers. Under 
degraded atmospheric conditions, flux dropouts will result in low S/N values. For low target altitudes, i.e. higher airmass, a trend is at least 
visible, which might be explained by the increased amount of turbulent atmospheric layers the light has to pass. Also, longitudinal dispersion 
increases, leading to a lower fringe contrast. A longitudinal atmospheric dispersion compensator (LADC) is present for the FSU-A \citep{sah09} but is 
not tested and considered here. In Figure~\ref{fig:lr7} we show plots for the FSU-A lock ratio as a function of the parameters as in 
Fig.~\ref{fig:gdres}. The lock ratio is defined as the number of frames with the fringe-tracking loop closed divided by the total number of frames 
expressed in percent. Correlations are less prominent but still reflect the general picture that the fringe-tracking performance decreases under 
degraded atmospheric conditions and low object altitude.\\

% \begin{figure}[!ht]
%   \centering
%   \includegraphics[width=9cm]{FSUA_Correlations_PDres.ps}
%   \caption{.}
%   \label{fig:pdres}
% \end{figure}

\subsection{Data reduction}

\begin{figure}[!ht]
  \centering
  \includegraphics[width=9cm]{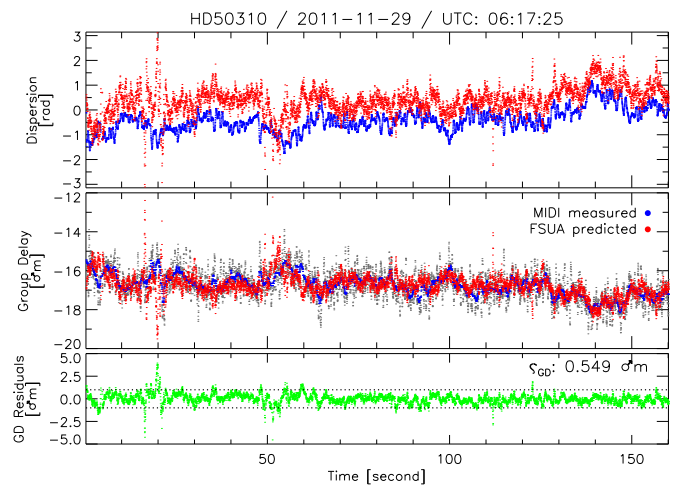}
  \caption{Upper panel: $N$ band dispersion measured by EWS on MIDI data (red points) and predicted values using the $K$ band data from FSU-A (blue points). Middle panel: measured (blue) and predicted (red) $N$ band group delay values. Lower panel: Group delay residual between measurement and prediction.}
  \label{fig:koresko}
\end{figure}

\begin{figure}[!ht]
  \centering
  \includegraphics[width=9cm]{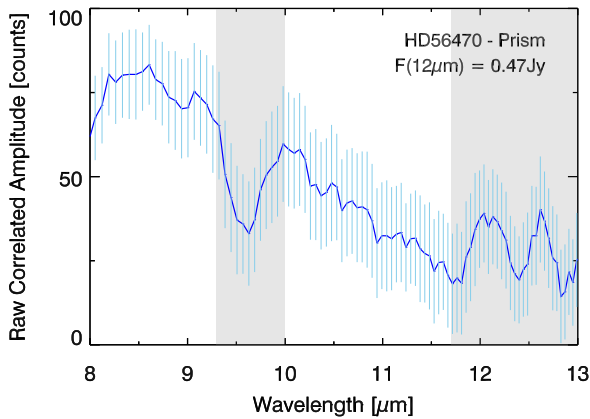}
  \caption{Raw correlated flux of star HD~56470 (IRAS flux density $F_N=0.47$\,Jy \citep{iras}) observed during a PRIMA commissioning run (December 12, 2009) using the ATs with the MIDI+FSU-A mode. The light blue vertical lines are the formal errors as produced by EWS. The strong ozone absorption feature between 9.3\,$\mu$m and 10.0\,$\mu$m is clearly visible in the plot and marked by the vertical gray region. For wavelengths beyond $\sim$11.7\,$\mu$m, the source flux of faint $N$ band targets drops to lower atmospheric transmission and cannot be measured at high accuracy because of high thermal background.}
  \label{fig:470mjy}
\end{figure}

\begin{figure}[!ht]
  \centering
  \includegraphics[width=9cm]{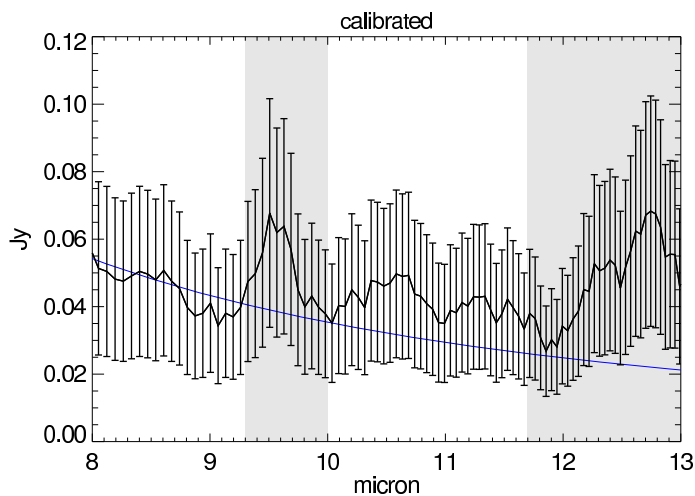}
  \caption{Same as Fig.~\ref{fig:470mjy} but for the $F_N=50$\,mJy source TYC~5282-2210-1 using the UTs. The blue line is a stellar photospheric fit 
to WISE \citep{wri10} photometry for an 
independent confirmation of the measured $N$ band flux density.}
  \label{fig:50mjy}
\end{figure}

The data reduction of bright $N$ band targets ($N\gtrsim5$\,Jy using the ATs) observed with the MIDI+FSU-A mode can be carried out using the standard 
data reduction routines of EWS\footnote[2]{The software package MIA+EWS (MIDI Interactive Analysis + Expert Work Station) is available for download 
at \url{http://www.strw.leidenuniv.nl/~nevec/MIDI/index.html}} \citep{jaf04a} and will not be explained further. For resolved or faint targets 
($N\lesssim5$\,Jy using the ATs), a different approach has to be chosen to reliably identify the MIDI interferograms and to measure the correlated 
fluxes. The $K$ band data of the FSU-A carry the information directly about whether at any given time a fringe was observed or not based on the 
status of the 
OPDC. Therefore, $N$ band data that are not directly visible in real time during the observation can be identified in post-processing by taking the 
FSU-A data into account. 
Using the HIGH\_SENS mode of MIDI (all light of the target is sent directly to the $N$ band beam combiner without recording photometry 
simultaneously), the integration time of one frame is 18\,ms. The FSU-A is usually operated with 1\,kHz leading to 18 FSU-A frames during one MIDI 
exposure. In a first step, the corresponding FSU-A frames for the individual MIDI frames thus have to be identified. A MIDI frame is considered 
useful if the OPDC state 
indicates a fringe lock for FSU-A for all 18 frames. To estimate the necessary $N$ band group delay and phase delay from $K$ band (the computation of 
the $K$ band phase and group delay is described in \citet{sah09}) data, we implemented the method of \citet{kor06}. They used $K$ band data to 
measure 
relative changes in the differential column density of water vapor. This is the main source of dispersion in the infrared regime, and its short 
time variability will lead to changes in the phase and group delay of the corresponding interferogram. Through linear 
relationships (Eqs.~9 to 14 in \citet{kor06}), \citet{kor06} estimated the $N$ band GD and PD from $K$ band data using the knowledge of the 
wavelength-dependent refractivity of water vapor molecules. The independently determined $N$ band GD and PD values are forwarded to EWS to finally 
integrate the MIDI fringes coherently. In Figure~\ref{fig:koresko} we show the result of the prediction of dispersion (upper panel) and 
group delay (middle panel) in the $N$ band from FSU-A $K$ band data. The source was a calibrator star (\object{HD~50310}) bright in 
$N$ band to determine these values independently with EWS for comparison.
The offset visible in dispersion between the measured and predicted values is explained by the zero point of the MIDI dispersion, which 
includes quasi-static dispersions of transmitted optics, neither seen by the FSU-A nor taken care of by the algorithm, which translates the measured 
$K$ band dispersion into $N$ band dispersion.
The resulting GD residuals between measured and predicted values are shown in the lower panel of 
Fig.~\ref{fig:koresko}. The standard deviation of the GD residuals is as low as 0.55\,$\mu$m (see also \citet{pot12}). The residuals for the 
dispersion are typically around 40\degr. For faint $N$ band sources, the predicted values are forwarded to the EWS pipeline allowing 
reliable correlated flux measurements of MIDI down to 500\,mJy (Fig.~\ref{fig:470mjy}) and 50\,mJy (Fig.~\ref{fig:50mjy}) in the case of ATs and UTs, 
respectively. For comparison, the official MIDI flux density limits as provided by ESO are $\approx$5\,Jy on the ATs and 0.2\,Jy on the UTs, 
respectively.

By using the AC-filtered FSU-data based predictions of MIDI group delay and dispersion we achieve artificially extended, 100 times 
longer $N$ band coherence times between 10 and 100 seconds, as opposed to 0.1--1 seconds for a MIDI stand-alone operation. These numbers are 
comparable to 
the performance of the KI-Nuller multiband phase and dispersion tracking, as described in \citet{col10}, who typically run the 
slow $N$ band control loop to identify offset drifts with cycles of coherent integration times of 1.25--6.25 seconds \citep{col10b}. The same 
timescale optimizes our $N$ band S/N. For particularly faint objects ($<$100\,mJy on the UTs), this coherent integration time scale can be increased 
to up to 60\,s, at the cost of precision and observing efficiency. Such long coherent integrations require longer overall integration times up to 
10 to 15 minutes. This length of total integration per $(u,\upsilon)$ point is a natural limit is set by the Earth's rotation (and respective change 
of 
$(u,\upsilon)$-coordinates at the scale of the interferometric resolution).

To obtain calibrated $N$ band visibilities, the correlated flux has to be divided by the measured total $N$ band flux (single-dish photometry). To obtain $N$ band photometry is the limiting factor of the MIDI visibility measurement. For sources fainter than 20\,Jy in the case of ATs and 1\,Jy for UTs, photometry can no longer be obtained by MIDI because of limitations in the background subtraction when chopping is applied \citep{bur13}. In addition, the relatively narrow field-of-view on the ATs hampers determining the sky background as well. Therefore, for such faint $N$ band targets, MIDI can only deliver precise correlated flux measurements. The precision of the visibility estimator in this $N$ band regime does not improve owing to FSU-A operation, because the total flux measurement is not helped by OPD stabilization. 
% Therefore, a comparison of MIDI $N$ band visibilities with and without the FSU-A cannot be carried out as the photometry is the main source of error.

% \missingfigure[figwidth=9cm]{Figure still has to be created.}
% \includegraphics[draft]{foo}

\begin{table*}
\caption{Observing log of MIDI+FSU-A observations using the ATs and UTs.\label{tbl:obsmf}}
%  The columns are: (1) the name of the target; (2) and (3) date and time of observation; (4) the AT/UT station used; (5) and (6) 
% $u,\upsilon$-coordinates; (7) and (8) length and position angle of the projected baseline; (9), (10), and (11) seeing conditions, airmass, 
% and coherence time $\tau_\text{0}$ at the time of observation.
\begin{tabular}{lcccccccccc}
\hline																		
				
\hline																		
				
(1)		&	(2)	&	(3)	&	(4)	&	(5)	&	(6)	&	(7)	&	(8)	&	(9)	
&	(10)	&	(11)	\\
Target		&	Date	&	Time / UT	&	Baseline name	&	$u$	&	$\upsilon$	&	$B_p$	&	
$PA$	&	Seeing	&	Airmass	&	$\tau_\text{0}$	\\
		&		&	[hh:mm]	&		&	[m]	&	[m]	&	[m]	&	[\degr]	
&	[\arcsec]	&		&	[ms]	\\
\hline																		
				
\multicolumn{11}{c}{ATs, program ID 087.C-0824}\\												
			
\object{HD 1014}	&	2011-11-28	&	00:59	&	A1-B2	&	-9.80	&	4.92	&	10.97	&	296.7	
&	0.71	&	1.05	&	5.91	\\
24 Psc	&	2011-11-28	&	01:18	&	A1-B2	&	-9.02	&	4.77	&	10.20	&	297.9	&	1.03	
&	1.13	&	4.07	\\
\object{HD 16212}	&	2011-11-28	&	02:21	&	A1-B2	&	-10.35	&	4.57	&	11.31	&	293.8	
&	0.85	&	1.05	&	4.84	\\
\object{NGC 1068}	&	2011-11-28	&	02:44	&	A1-B2	&	-10.26	&	4.50	&	11.20	&	293.7	
&	1.08	&	1.10	&	5.39	\\
HD 16212	&	2011-11-28	&	03:06	&	A1-B2	&	-10.00	&	4.84	&	11.10	&	295.8	&	
1.08	&	1.05	&	3.56	\\
HD 1014	&	2011-11-28	&	03:30	&	A1-B2	&	-5.57	&	5.64	&	7.93	&	315.4	&	1.64	
&	1.45	&	2.35	\\
24 Psc	&	2011-11-28	&	04:04	&	A1-B2	&	-3.32	&	5.03	&	6.03	&	326.6	&	1.46	
&	2.05	&	2.65	\\
\object{HD 28305}	&	2011-11-28	&	04:33	&	A1-B2	&	-10.24	&	3.69	&	10.89	&	289.8	
&	1.39	&	1.39	&	2.83	\\
HD 16212	&	2011-11-28	&	05:18	&	A1-B2	&	-6.84	&	5.52	&	8.79	&	308.9	&	
0.81	&	1.29	&	4.80	\\
\object{HD 36167}	&	2011-11-28	&	06:25	&	A1-B2	&	-9.61	&	4.57	&	10.64	&	295.4	
&	0.69	&	1.11	&	5.69	\\
\object{$\delta$ Ori~A}	&	2011-11-28	&	06:44	&	A1-B2	&	-9.26	&	4.52	&	10.31	&	296.0	
&	0.80	&	1.13	&	4.84	\\
HD 36167	&	2011-11-28	&	07:05	&	A1-B2	&	-8.77	&	4.60	&	9.91	&	297.7	&	
0.96	&	1.16	&	3.98	\\
$\delta$ Ori~A	&	2011-11-28	&	07:16	&	A1-B2	&	-8.53	&	4.53	&	9.66	&	298.0	&	
1.10	&	1.19	&	3.49	\\
\object{HD 50778}	&	2011-11-28	&	07:31	&	A1-B2	&	-9.89	&	5.06	&	11.11	&	297.1	
&	1.01	&	1.03	&	3.77	\\
\object{HD 53179}	&	2011-11-28	&	07:58	&	A1-B2	&	-9.61	&	5.19	&	10.92	&	298.4	
&	1.81	&	1.04	&	2.72	\\
24 Psc	&	2011-11-29	&	02:31	&	A1-D0	&	-26.36	&	-22.81	&	34.86	&	229.1	&	1.19	
&	1.34	&	3.21	\\
HD 1014	&	2011-11-29	&	02:51	&	A1-D0	&	-26.44	&	-22.13	&	34.48	&	230.1	&	1.09	
&	1.28	&	3.15	\\
HD 16212	&	2011-11-29	&	03:12	&	A1-D0	&	-26.13	&	-24.04	&	35.51	&	227.4	&	
1.21	&	1.05	&	2.83	\\
\object{HD 27639}	&	2011-11-29	&	03:53	&	A1-D0	&	-23.04	&	-16.55	&	28.37	&	234.3	
&	1.38	&	1.45	&	2.45	\\
HD 27639	&	2011-11-29	&	04:12	&	A1-D0	&	-24.09	&	-17.19	&	29.59	&	234.5	&	
1.25	&	1.43	&	2.69	\\
\object{RY Tau}	&	2011-11-29	&	04:55	&	A1-D0	&	-26.10	&	-16.73	&	31.00	&	237.4	&	
1.12	&	1.67	&	2.95	\\
HD 50778	&	2011-11-29	&	05:36	&	A1-D0	&	-19.31	&	-26.71	&	32.96	&	215.9	&	
1.11	&	1.10	&	2.93	\\
HD 53179	&	2011-11-29	&	05:57	&	A1-D0	&	-20.35	&	-26.37	&	33.31	&	217.7	&	
0.68	&	1.08	&	4.83	\\
\object{HD 50310}	&	2011-11-29	&	06:17	&	A1-D0	&	-22.72	&	-26.32	&	34.77	&	220.8	
&	0.95	&	1.13	&	3.37	\\
\object{$\beta$ Pic}	&	2011-11-29	&	06:30	&	A1-D0	&	-26.43	&	-19.67	&	32.95	&	233.4	
&	0.80	&	1.12	&	4.00	\\
HD 50310	&	2011-11-29	&	06:46	&	A1-D0	&	-24.44	&	-24.00	&	34.26	&	225.5	&	
0.77	&	1.11	&	4.10	\\
\object{HD 53047}	&	2011-11-29	&	07:01	&	A1-D0	&	-24.65	&	-23.51	&	34.07	&	226.4	
&	0.96	&	1.12	&	3.30	\\
\object{HD 83618}	&	2011-11-29	&	07:44	&	A1-D0	&	-15.99	&	-23.65	&	28.55	&	214.1	
&	0.83	&	1.27	&	3.73	\\
\object{HD 92305}	&	2011-11-29	&	08:28	&	A1-D0	&	-14.82	&	-26.97	&	30.78	&	208.8	
&	1.04	&	1.79	&	2.95	\\
HD 1014	&	2011-11-30	&	00:46	&	B2-D0	&	-16.25	&	-28.94	&	33.19	&	209.3	&	2.02	
&	1.05	&	1.84	\\
\object{HD 1522}	&	2011-11-30	&	01:26	&	B2-D0	&	-17.72	&	-28.66	&	33.70	&	211.7	
&	1.53	&	1.07	&	2.50	\\
24 Psc	&	2011-11-30	&	01:46	&	B2-D0	&	-19.05	&	-27.92	&	33.80	&	214.3	&	1.78	
&	1.20	&	2.10	\\
HD 1014	&	2011-11-30	&	02:03	&	B2-D0	&	-18.94	&	-28.13	&	33.91	&	214.0	&	1.28	
&	1.15	&	2.95	\\
24 Psc	&	2011-11-30	&	02:24	&	B2-D0	&	-19.53	&	-27.75	&	33.93	&	215.1	&	1.11	
&	1.32	&	3.40	\\
HD 1522	&	2011-11-30	&	02:46	&	B2-D0	&	-19.49	&	-27.66	&	33.84	&	215.2	&	1.12	
&	1.25	&	4.23	\\
24 Psc	&	2011-11-30	&	03:04	&	B2-D0	&	-19.45	&	-27.56	&	33.73	&	215.2	&	1.37	
&	1.54	&	3.41	\\
HD 27639	&	2011-11-30	&	03:43	&	B2-D0	&	-12.14	&	-20.45	&	23.78	&	210.7	&	
0.93	&	1.46	&	4.93	\\
RY Tau	&	2011-11-30	&	04:03	&	B2-D0	&	-13.48	&	-17.61	&	22.18	&	217.4	&	0.73	
&	1.68	&	6.19	\\
HD 16212	&	2011-11-30	&	04:21	&	B2-D0	&	-18.86	&	-28.17	&	33.90	&	213.8	&	
0.73	&	1.14	&	6.00	\\
HD 50778	&	2011-11-30	&	06:39	&	B2-D0	&	-13.82	&	-30.00	&	33.03	&	204.7	&	
0.90	&	1.03	&	4.69	\\
HD 53179	&	2011-11-30	&	07:11	&	B2-D0	&	-15.10	&	-29.65	&	33.27	&	207.0	&	
0.72	&	1.03	&	5.78	\\
HD 92305	&	2011-11-30	&	08:11	&	A1-D0	&	-13.51	&	-27.76	&	30.87	&	206.0	&	
0.89	&	1.81	&	4.72	\\
\object{HD 100546}	&	2011-11-30	&	08:48	&	A1-D0	&	-12.25	&	-30.73	&	33.09	&	201.8	
&	0.98	&	1.60	&	4.36	\\
\hline
\multicolumn{11}{c}{UTs, program ID 090.B-0938}\\												
			
\object{HD 10380}	&	2012-11-04	&	02:55	&	UT1-UT2	&	 21.49	&	 43.68	&	48.88	&	 26.7	
&	0.84	&	1.17	&	2.66	\\
\object{TYC 5282-2210-1}	&	2012-11-04	&	03:58	&	UT1-UT2	&	 24.50	&	 49.34	&	55.09	&	 
26.4	&	1.37	&	1.03	&	1.66	\\

\hline
\end{tabular}
%   \tablefoot{The columns are: (1) index number by which an observation can be identified in tables and figures; (2) the name of the observed 
% object; (3) date;\\ (4) time; (5) telescope stations; (6) and (7) $u,v$-coordinates; (8) length of the projected baseline; (9) position angle of the 
% projected baseline; (10) and\\ (11) model visibility in $K$ band for FSU-A and in $N$ band for MIDI, respectively; (12) FSU-A operation frequency; 
% (13) IRIS DIT; (14) airmass;\\ (15) seeing; (16) coherence time; (17) relative humidity.\\}
\end{table*}

%______________________________________________________________

\section{$K$ band visibility estimate}\label{sec:vis}

With the MIDI+FSU-A mode, it is possible to scan through the fringe package using the main DLs and record the temporal encoded interferometric signal with the FSU-A before the actual MIDI+FSU-A observation starts. With a typical setup of 300 scans over an OPD range of 350\,$\mu$m, this observation takes less than four minutes producing only a marginal overhead. The scanning technique allows the measurements of calibrated visibilities for scientific targets in the $K$ band, and thus complements the $N$ band observations, providing a more complete picture of the source of interest. In this section we present first observations obtained with the scanning technique.

\subsection{Observations and data reduction}\label{sec:visobsdr}

To validate the visibility measurements from FSU-A data we observed two binary stars \object{HD~155826} and \object{24~Psc}\footnote[3]{All presented 
scanning data are publicly available for download from the ESO archive at \url{http://archive.eso.org/eso/eso_archive_main.html}. They can be 
retrieved by entering the corresponding date of the observations and by selecting ``MIDI/VLTI'' as instrument.}. Both stars have well known orbits 
\citep{mas10} and are therefore well suited for testing our algorithm to compare measurements versus predicted visibility values. The orbital 
elements of 
the two binary systems are presented in Table~\ref{tbl:sci} and were originally published by \citet{mas10}.
We were using the following instrumental setup: the FSU-A was operated at 1\,kHz (which corresponds to an integration time of $\approx$1\,ms per 
frame), a scan range of 350\,$\mu$m was used, and at a scanning speed of $\approx$550\,$\mu$m\,$s^{-1}$, the resulting sampling rate was two samples 
per 
micrometer, which is close to the Nyquist frequency. The observing sequence was chosen in such a way that the science target were bracketed by two 
different calibrators, i.e. Cal1--Sci--Cal2. HD~155826 was observed seven times in the night of July 6, 2013 over the course of three hours on a 
36\,m baseline (AT stations A1 and D0). The star 24~Psc was observed five times in the night of October 28, 2013 over the course of two hours on a 
80\,m baseline (AT stations A1 and G1). Table~\ref{tbl:obs} lists the individual observations and which conditions they were taken under.\\

\begin{table}[!ht]
  \caption{Orbital elements (period $P$, semimajor axis $a$, inclination $i$, longitude of the ascending node $\varOmega$, epoch of periastron 
passage $T_{\text{0}}$, eccentricity $e$, longitude of periastron $\omega$) of HD~155826 and 24~Psc as published by \citet{mas10}. \label{tbl:sci}}
  \centering
  \begin{tabular}{lcc}
  \hline\hline
  Parameter & HD~155826  & 24~Psc \\
  \hline
  $K$ / [mag]           & 4.555$^{a}$ &  3.684$^{a}$\\
  $P$ / [yr]            & 14.215  & 22.81 \\
  $a$ / [\arcsec]       & 0.2527  & 0.0832 \\
  $i$ / [\degr]    & 115.2  & 133.7 \\
  $\varOmega$ / [\degr] & 190.41  & 209.5 \\
  $T_{\text{0}}$ / [yr] & 1985.98  & 1988.72 \\
  $e$                   & 0.4912  & 0.422 \\
  $\omega$ [\degr] & 135.2  & 298.3 \\
  flux ratio            & 4.3$^{b}$ & 1.0$^{b}$\\
  $\rho$ / [\arcsec]    & 0.075$^{c}$ & 0.063$^{c}$\\
  $\theta$ / [\degr] & 133.1$^{c}$ & 195.0$^{c}$\\
  \hline
  \end{tabular}
    \tablefoot{$^{(a)}$ SIMBAD magnitudes. $^{(b)}$ Magnitudes of the individual components were obtained from the Washington Double Star Catalog. 
$^{(c)}$ The values for the apparent separation $\rho$ and position angle $\theta$ of the companions as computed for the date of observation.}
\end{table}

\begin{table*}[!ht]
  \caption{Observing log of FSU-A fringe scan observations. The stellar diameters $\theta$ of the used calibrator stars were taken from the MIDI 
calibrator database provided through R. van Boekel which is distributed by the MIDI software reduction package MIA+EWS.\label{tbl:obs}}
%  The columns are: (1) the name of the target; (2) time of observation in UT; (3) and (4) $u,\upsilon$-coordinates; (5) and (6) length and position 
% angle of the projected baseline; (7), (8), and (9) seeing conditions, object altitude, and coherence time $\tau_\text{0}$ at the time of observation; % (10) science or calibrator observation.
  \centering
  \begin{tabular}{lccccccccl}
  \hline\hline
  (1)    & (2)        & (3) & (4) & (5)   & (6)           & (7)          & (8)          & (9)             &      (10)     \\
  Target & Time / UT  & $u$ & $\upsilon$ & $B_p$  &  $PA$         & Seeing       & Altitude     & $\tau_\text{0}$ &      Comments \\
         & [hh:mm:ss] & [m] & [m] & [m]   & [\degr]  & [\arcsec]    & [\degr] & [ms]            &                   \\
  \hline
 \multicolumn{10}{c}{Observations of HD~155826 in the night of July 6, 2013}\\									
						
HD152161	&	02:01:52	&	-23.42	&	-26.44	&	35.32	&	221.54	&	0.80	&	70.54	&	
2.99	&	calibrator, $\theta=6.036$mas	\\
HD 155826	&	02:17:12	&	-23.02	&	-27.22	&	35.65	&	220.23	&	0.82	&	73.98	&	
2.91	&		\\
HD152334	&	02:32:34	&	-25.03	&	-24.37	&	34.94	&	225.76	&	0.87	&	72.27	&	
2.83	&	calibrator, $\theta=4.107$mas	\\
HD 155826	&	02:44:01	&	-24.56	&	-25.47	&	35.39	&	223.96	&	0.74	&	75.87	&	
3.28	&		\\
HD152161	&	02:56:34	&	-26.04	&	-22.38	&	34.34	&	229.33	&	0.83	&	70.91	&	
2.94	&	calibrator, $\theta=6.036$mas	\\
HD 155826	&	03:08:56	&	-25.69	&	-23.76	&	35.00	&	227.24	&	0.84	&	75.69	&	
2.90	&		\\
HD152334	&	03:21:59	&	-26.74	&	-20.59	&	33.75	&	232.41	&	0.80	&	69.60	&	
3.07	&	calibrator, $\theta=4.107$mas	\\
HD 155826	&	03:35:47	&	-26.57	&	-21.84	&	34.40	&	230.58	&	0.76	&	73.48	&	
3.24	&		\\
HD152161	&	03:47:39	&	-27.15	&	-18.30	&	32.75	&	236.02	&	0.72	&	65.93	&	
3.43	&	calibrator, $\theta=6.036$mas	\\
HD 155826	&	04:00:27	&	-27.06	&	-20.04	&	33.67	&	233.48	&	0.79	&	70.18	&	
3.14	&		\\
HD152334	&	04:21:44	&	-27.15	&	-15.81	&	31.42	&	239.78	&	0.74	&	61.43	&	
3.39	&	calibrator, $\theta=4.107$mas	\\
HD 155826	&	04:34:57	&	-27.20	&	-17.48	&	32.33	&	237.28	&	0.79	&	64.49	&	
3.17	&		\\
HD152161	&	04:47:14	&	-26.74	&	-13.48	&	29.95	&	243.25	&	0.76	&	56.76	&	
3.30	&	calibrator, $\theta=6.036$mas	\\
HD 155826	&	04:58:57	&	-26.94	&	-15.70	&	31.18	&	239.76	&	0.81	&	60.13	&	
3.06	&		\\
HD152334	&	05:11:09	&	-26.09	&	-11.92	&	28.69	&	245.45	&	0.82	&	52.98	&	
3.05	&	calibrator, $\theta=4.107$mas	\\
\hline																		
	
\multicolumn{10}{c}{Observations of 24~Psc in the night of October 28, 2013}\\									
										
HD28	&	04:39:38	&	-55.23	&	27.52	&	61.71	&	296.49	&	0.80	&	50.34	&	2.79	
&	calibrator, $\theta=1.730$mas	\\
24 Psc	&	04:51:23	&	-49.30	&	25.53	&	55.52	&	297.38	&	0.78	&	43.88	&	2.79	
&		\\
HD223800	&	05:12:56	&	-43.50	&	35.02	&	55.84	&	308.84	&	0.65	&	43.12	&	
3.06	&	calibrator, $\theta=3.559$mas	\\
24 Psc	&	05:23:16	&	-40.63	&	25.88	&	48.17	&	302.5	&	0.71	&	36.98	&	2.79	
&		\\
HD28	&	05:35:18	&	-40.73	&	28.69	&	49.82	&	305.16	&	0.70	&	38.25	&	2.82	
&	calibrator, $\theta=1.730$mas	\\
24 Psc	&	05:46:28	&	-33.81	&	26.09	&	42.71	&	307.65	&	0.78	&	31.85	&	2.52	
&		\\
HD223800	&	05:57:52	&	-30.29	&	36.53	&	47.46	&	320.34	&	0.69	&	32.91	&	
2.89	&	calibrator, $\theta=3.559$mas	\\
24 Psc	&	06:08:12	&	-27.11	&	26.25	&	37.74	&	314.07	&	0.95	&	26.99	&	2.07	
&		\\
HD28	&	06:19:52	&	-27.34	&	29.35	&	40.11	&	317.03	&	0.94	&	28.25	&	2.10	
&	calibrator, $\theta=1.730$mas	\\
24 Psc	&	06:30:31	&	-19.98	&	26.37	&	33.08	&	322.86	&	0.93	&	21.95	&	2.12	
&		\\
HD28	&	06:41:16	&	-20.50	&	29.58	&	35.99	&	325.27	&	0.96	&	23.39	&	2.06	
&	calibrator, $\theta=1.730$mas	\\

  \hline
  \end{tabular}
\tablefoot{The software package “MIA+EWS” is available for download at \url{http://www.strw.leidenuniv.nl/~nevec/MIDI/index.html}.}
  \end{table*}

\begin{figure}[!ht]
  \centering
  \includegraphics[width=9cm]{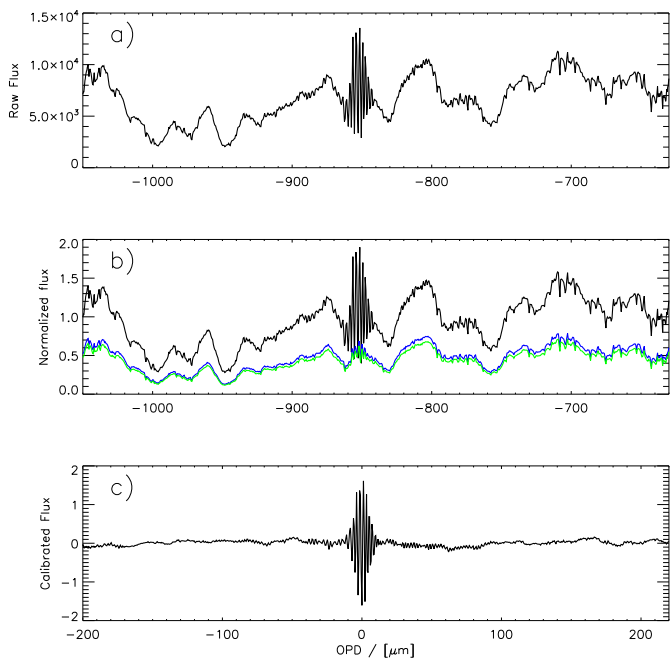}
  \caption{Illustration of the individual reduction steps for a synthetic white light pixel. a) Raw fringe scan. b) Normalized fringe scan; i.e., 
flat field and sky background correction were applied. Superimposed in blue and green are the extracted photometries of telescope beam T1 and T2. c) 
Photometric calibrated and normalized interferogram. See text for explanation.}
  \label{fig:fringe}
\end{figure}

The raw fluxes (Fig.~\ref{fig:fringe}a) are corrected for sky background and by flat fields (see Eq.2 in \citet{sah09}) and will be denoted as 
$IA_{n,j}$, $IB_{n,j}$, $IC_{n,j}$, and $ID_{n,j}$ in the following, where $n$ is the scan number, and $j$ denotes the corresponding spectral pixel 
(0 to 5, where 0 stands for the synthetic white light pixel). Since the flat fields are recorded on the target itself at each new preset, we can use 
them to determine an internal flux splitting ratio (similar to a $\kappa$ matrix) even though the FSU-A does not have separate photometric channels 
to monitor the photometry of the individual beams. To do that, we divide the flat fields of the two beams:
\begin{equation}
  \kappa^{\text{T2}}_{i,j} = \overline{F^{\text{T2}}_{i,j}}/\overline{F^{\text{T1}}_{i,j}},
\end{equation}
where $\overline{F}$ is the average flux value measured by the flat field for telescope beam T1 and T2, respectively. The subscripts $i$ denote the 
corresponding channel A, B, C, or D, and $j$ denotes the corresponding spectral pixel. Because we have only two beams, the flux splitting ratio 
cannot be determined in a unique way (in contrast to PIONIER, which operates with four telescopes and a pairwise $\kappa$ matrix can be computed 
\citep{leb11}); i.e., we set the beam splitting ratio for beam 1 to one with $\kappa^{\text{T1}}_{\textrm{i,j}}=1$. 
Real-time photometry is required for normalization and calibration purposes of the fringe package. To extract the real-time photometry obtained 
during 
the scan, we assume that the flux injection into all four channels is correlated. In addition, based on the ABCD algorithm, channel C has a spacial 
phase shift of 180\degr~with respect to channel A, as well as channel D with respect to channel B. Using this property, we can sum up channels A and 
C, and B and D. The fringe package disappears, and only the photometric signal remains (Fig.~\ref{fig:fringe}b).  
\begin{equation}
  \begin{array}{l}
  \displaystyle PA^{\text{T1}}_{n,j} = (IA_{n,j}+IC_{n,j})/4\\
  \displaystyle PA^{\text{T2}}_{n,j} = \kappa^{\text{T2}}_{\textrm{A},j} \cdot (IA_{n,j}+IC_{n,j})/4\\
  \displaystyle PC^{\text{T1}}_{n,j} = (IC_{n,j}+IA_{n,j})/4\\
  \displaystyle PC^{\text{T2}}_{n,j} = \kappa^{\text{T2}}_{\textrm{C},j} \cdot (IC_{n,j}+IA_{n,j})/4\\
  \vdots
  \end{array} 
  \label{eqn:phot}
\end{equation}
where $PA$ is the extracted photometric signal for channel A and for telescope beams T1 and T2. The division by four arises from the fact that we sum 
up two channels with flux contributions of two beams. The computation for $PB$ and $PD$ are similar. For the photometric calibration of the fringe 
signal, we use the definition by \citet{mer06}:
\begin{equation}
  \begin{array}{l}
  \displaystyle A_{n,j} = 
\frac{1}{\sqrt{\kappa^{\text{T2}}_{\textrm{A},j}}}\frac{IA_{n,j}-PA^{\text{T1}}_{n,j}-\kappa^{\text{T2}}_{\textrm{A},j}PA^{\text{T2}}_{n,j}}{\sqrt{
\overline{PA^{\text{T1}}_{n,j}}\cdot \overline{PA^{\text{T2}}_{n,j}}}}\\
  \displaystyle B_{n,j} = 
\frac{1}{\sqrt{\kappa^{\text{T2}}_{\textrm{B},j}}}\frac{IB_{n,j}-PB^{\text{T1}}_{n,j}-\kappa^{\text{T2}}_{\textrm{B},j}PB^{\text{T2}}_{n,j}}{\sqrt{
\overline{PB^{\text{T1}}_{n,j}}\cdot \overline{PB^{\text{T2}}_{n,j}}}}\\
  \vdots
  \end{array} 
  \label{eqn:cal}
\end{equation}
where $A_{n,j}$ and $B_{n,j}$ are the photometric calibrated fringe scans (Fig.~\ref{fig:fringe}c). Remember that $\kappa^{\text{T1}}_{i,j}$ was set 
to 
1. The computation for $C_{n,j}$ and $D_{n,j}$ is similar. The way the real-time photometry was extracted for the scans (summing up the channels A+C 
and B+D, Eq.\ref{eqn:phot}) results in mirror symmetric photometric residuals. Therefore, a further subtraction of the channel C from A, and D from 
B, respectively, results in the same signal; that is, because of the 180\degr~phase shift between A and C, $A_{n,j}-C_{n,j}$ would result in 
$A_{n,j}$ 
again. The same applies for the signals of channels B and D. Therefore, we denote all quantities as ``AC'' and ``BD'' in the following. The 
individual scans of the interferograms are now fully reduced and photometrically calibrated and can be used for visibility measurements.

\subsection{Fringe scan analysis}

\begin{figure}
\centering
\includegraphics[angle=0, width=9cm]{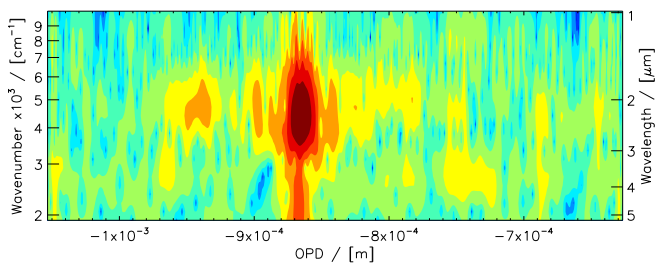}
\caption{Contour plot of a wavelet transform of the white light pixel of a single scan. The redder the color, the higher the wavelet power. The 
contours are plotted on a logarithmic scale.}
\label{fig:wavelet}
\end{figure}

\begin{figure*}
\centering
\includegraphics[angle=0, width=18cm]{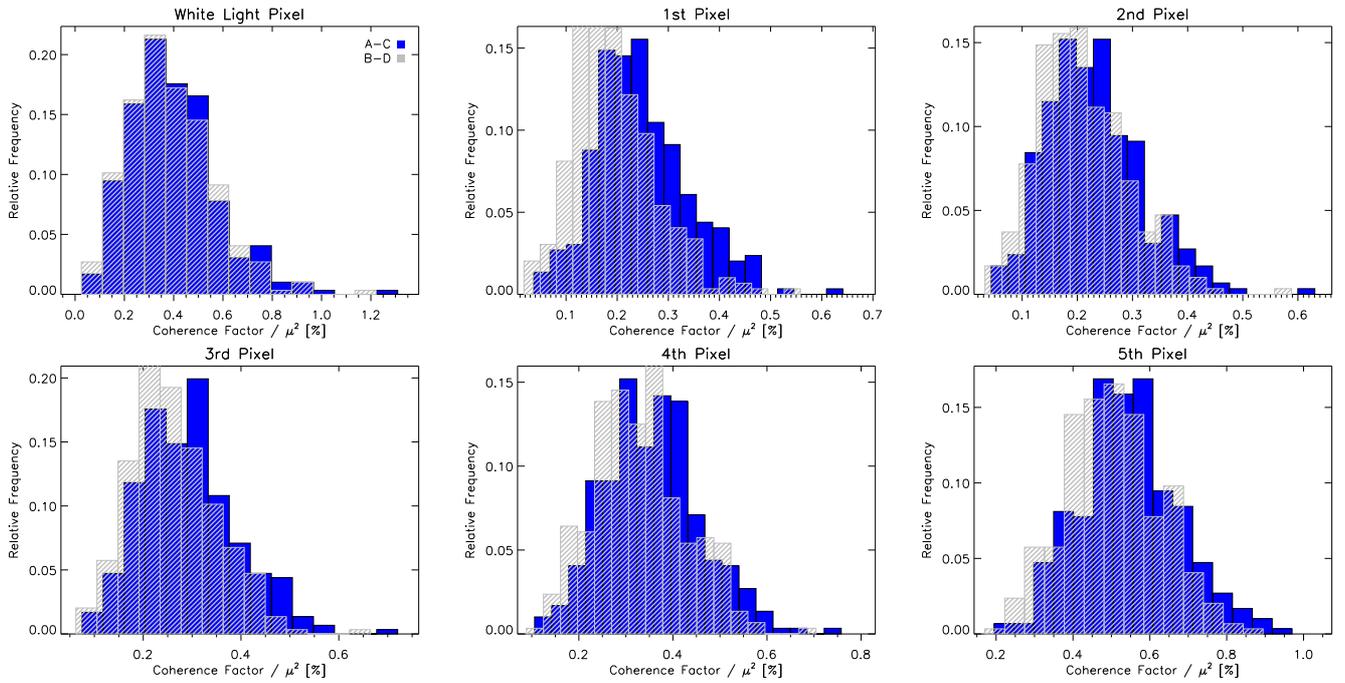}
\caption{Histograms of measured $\mu^2$ values obtained from a scan of the used calibrator star HD~152161 for all pixels of ``AC'' (in blue) and 
``BD'' (in gray).}
\label{fig:hist}
\end{figure*}

The analysis process of the interferograms of the fringe scans is mainly based on the method described in \citet{ker04} for VINCI data. For each 
interferogram, a wavelet transform is computed\footnote[4]{Wavelet software was provided by C. Torrence and G. Compo, and is available at 
\url{http://atoc.colorado.edu/research/wavelets/}.} and the used wavelet is a Morlet function. A wavelet transform has the advantage of 
simultaneously measuring the 
fringe position (zero OPD offset) and its frequency (i.e., wavelength). Figure~\ref{fig:wavelet} shows a wavelet transform of the 
white light pixel of a single scan. Because the synthetic white light pixel yields the highest S/N, we use it to determine the zero 
OPD offset in each scan and correct for it; i.e., all fringes are centered on zero OPD after the correction. In a second step, we compute a 
wavelet transform based on a theoretical interferogram that was used as a quality check for the real data.\\
As described in \citet{ker04}, we use 
three parameters to check the quality of the fringe packages. Scans were rejected if (1) the peak position in frequency domain deviates more than 
30\%, (2) the full width at half maximum (FWHM) exceeds 40\%, and (3) if the FWHM in OPD domain exceeds 50\% of that of the reference signal. The 
quality check is performed on only the synthetic white light pixel. Finally, the wavelet transforms of the 
remaining fringe scans are integrated over OPD and frequency (wavenumber) to obtain the coherence factor $\mu^2$. After the integration over OPD the 
now one-dimensional signal was fitted by a Gauss 
function with a third-degree polynomial function to account for bias power present in the data. As an example, in Figure~\ref{fig:hist} we show the 
measured $\mu^2$ values of a calibrator observation 
(HD~152161) for all spectral pixels (including the synthetic white light pixel). The histograms for ``AC'' and ``BD'' are superposed and symmetric, 
which is one characteristic to check for in case of 
present systematics or biases. Averaged 
coherence factors $\mu^2$ were computed from the histograms through bootstrapping.

\subsection{Visibility calibration}

\begin{figure*}
\centering
\includegraphics[angle=0, width=18cm]{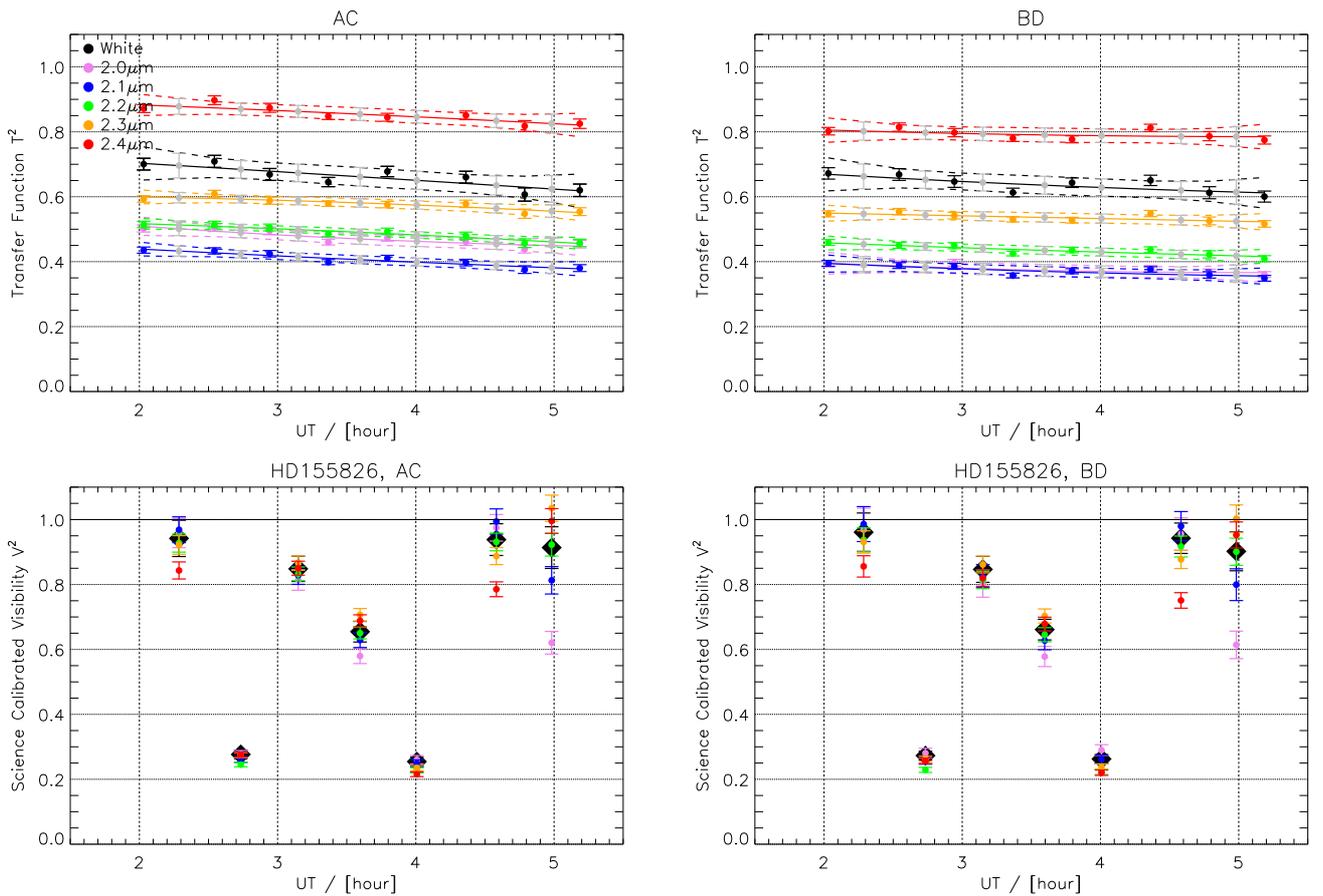}
\caption{Upper plots: Transfer functions $T^2$ for ``AC'' and ``BD'' for each spectral pixel and calibrator observation. The values for the 
individual spectral pixels are color-coded with respect to their corresponding wavelength (see legend in the upper left). The dashed lines are the 
corresponding 95\% confidence intervals of the polynomial fits shown as solid lines. They gray data points are the computed $T^2$ values for the time 
of the science observation.  Lower plots: Calibrated visibilities for each spectral pixel}
\label{fig:tf}
\end{figure*}

The construction of the transfer function $T^2$ for ``AC'' and ``BD'' from calibrator observations is straightforward and given by
\begin{equation}
 T^2 = \frac{\mu^2_{\text{cal}}}{V^2_{\text{cal}}},
\end{equation}
where $V^2_{\text{cal}}$ is the theoretical visibility for a given baseline (see Eq.\ref{eqn:v2}). The theoretical visibility was computed by using 
the uniform disk model:
\begin{equation}
 V^2_{\text{cal}} = \left| 2\frac{J_{1}(\pi \theta \nu)}{\pi \theta \nu}\right|,
 \label{eqn:v2}
\end{equation}
where $J_{1}$ is the Bessel function of the first kind, $\theta$ the stellar diameter (see Table~\ref{tbl:obs}), and $\nu$ the spatial frequency. 
The transfer functions for ``AC'' and ``BD'' for each spectral pixel and calibrator observation are displayed in the upper two plots of 
Figure~\ref{fig:tf}. For each spectral pixel the $T^2$ was fitted by a second-degree polynomial, which was used to compute the value of $T^2$ at the 
time of the science observation. The calibrated visibility $V^2_{\text{sci}}$ for the science observations can be derived by computing
\begin{equation}
 V^2_{\text{sci}} = \frac{\mu^2_{\text{sci}}}{T^2}.
\end{equation}
The lower plots of Figure~\ref{fig:tf} show the measured visibilities of the observations of the binary star HD~155826 for all spectral channels for 
``AC'' and ``BD'', respectively. Strong variations are evident caused by the change in baseline orientation during the course of the observation.

\subsection{Analysis of visibility measurements}

\begin{figure}[!ht]
  \centering
  \includegraphics[width=9cm]{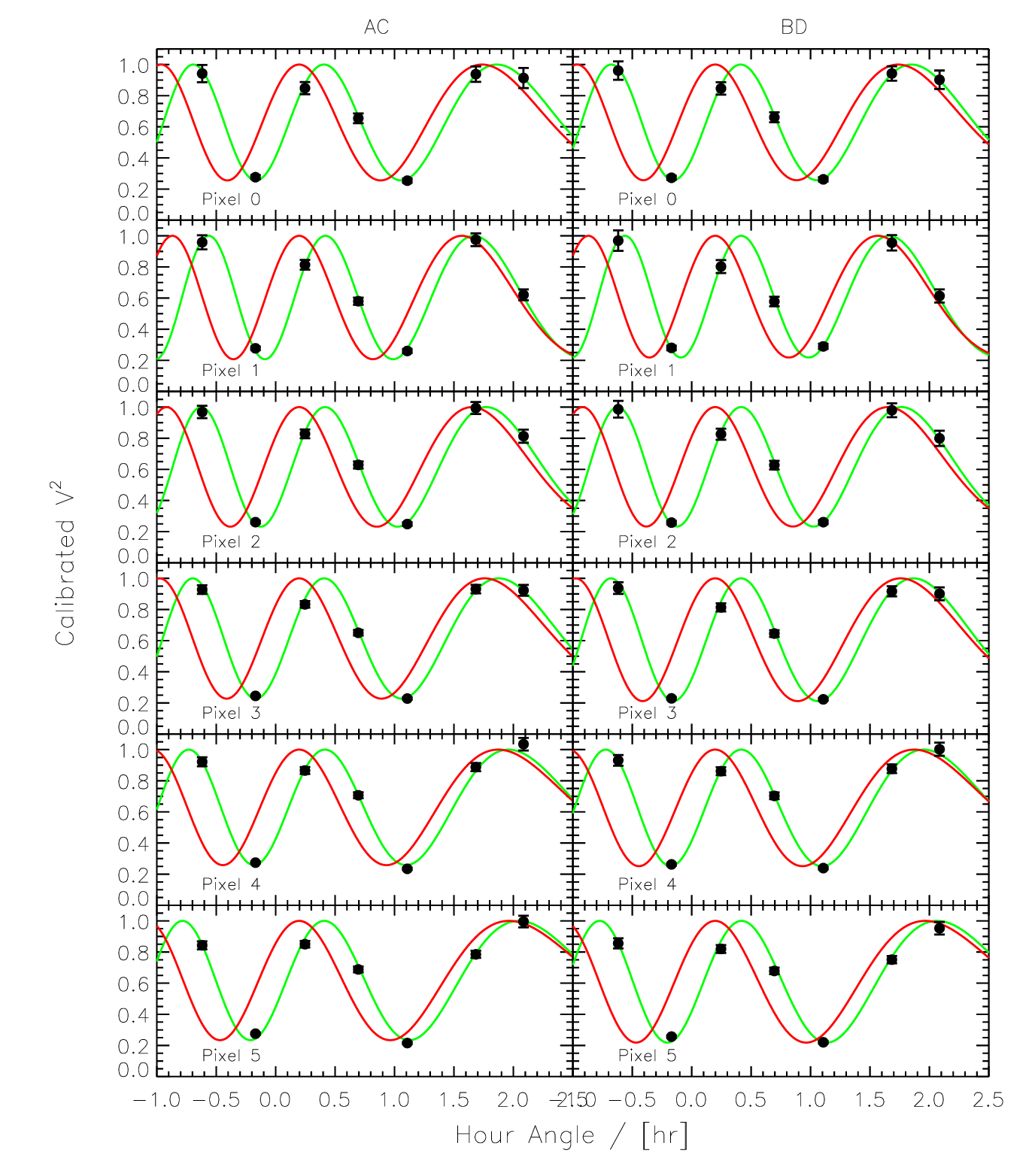}
  \caption{Fit of a binary model (green line) to the individual FSU-A visibility measurements (black points) for all spectral pixel of ``AC'' and 
``BD'' for HD~155826. The red line is the predicted visibility variation based on the published orbit by \citet{mas10}.}
  \label{fig:binmodelfit}
\end{figure}

\begin{figure}[!ht]
  \centering
  \includegraphics[width=9cm]{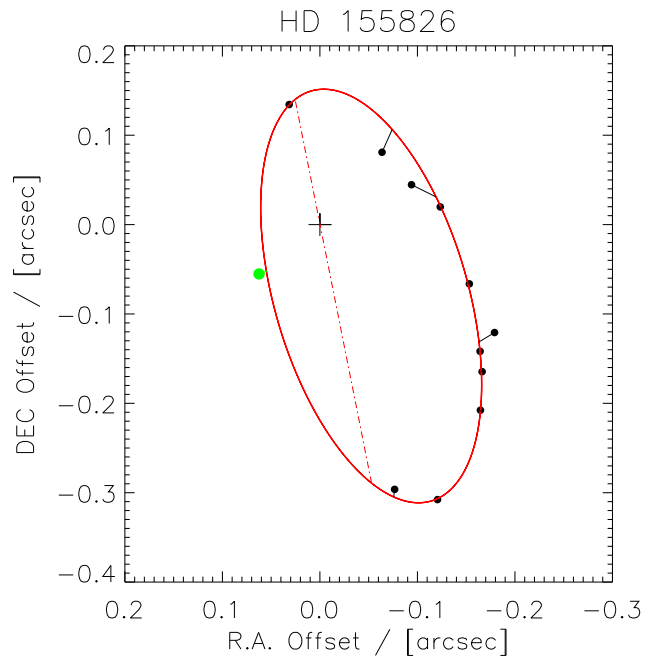}
  \caption{Astrometric orbit by \citet{mas10} of HD~155826. The published orbit (see Table~\ref{tbl:sci}) is shown as a red solid line, the dash 
dotted line is the line of nodes. The black points are individual measurements retrieved from the Washington Double Star Catalog. The green point is 
the new astrometric point based on FSU-A visibility measurements. The solid lines connecting the individual measurements with the orbit indicate the 
expected position for the given orbital solution.}
  \label{fig:hd155826}
\end{figure}

\begin{figure}[!ht]
  \centering
  \includegraphics[width=9cm]{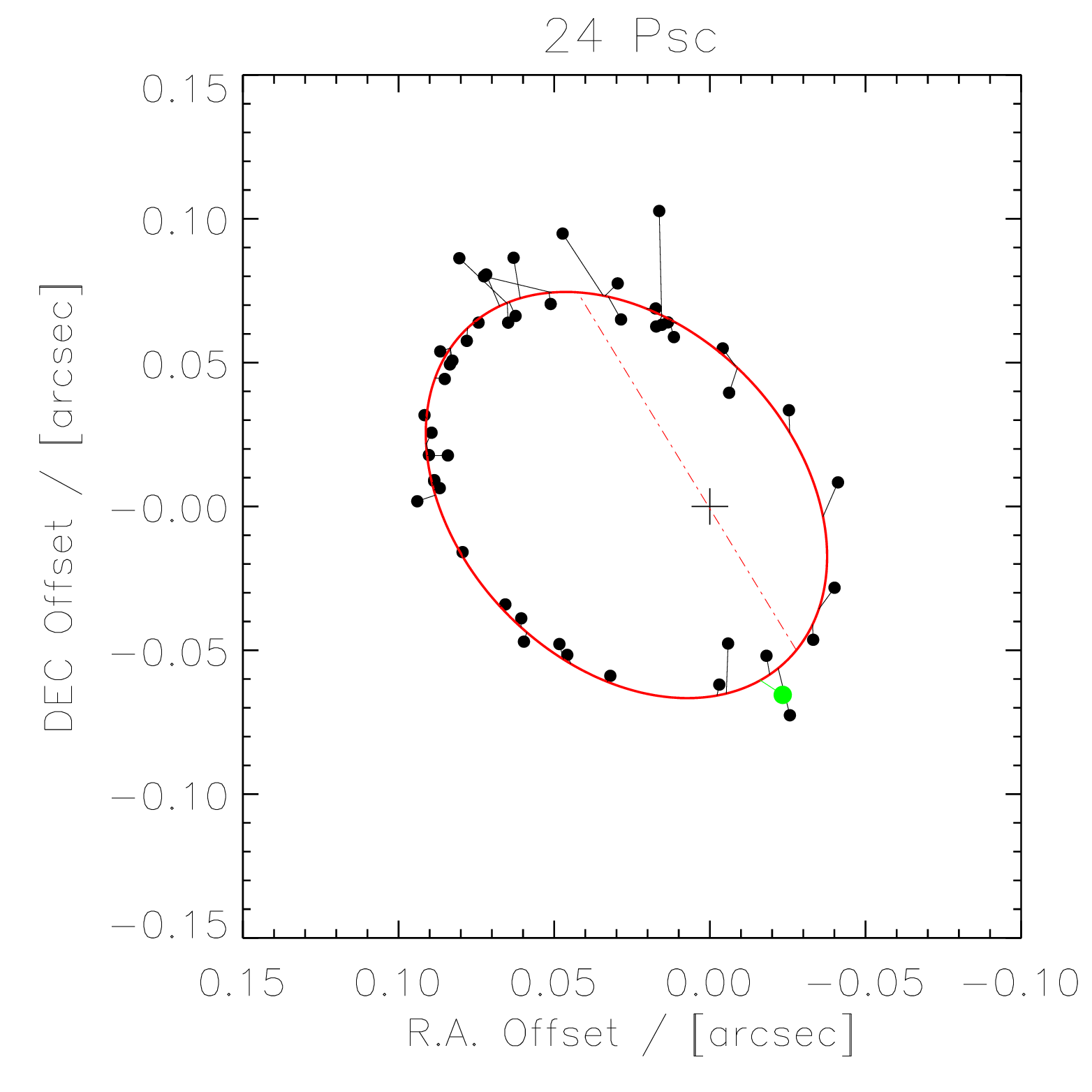}
  \caption{Same as Fig.~\ref{fig:hd155826} but for 24~Psc.}
  \label{fig:24psc}
\end{figure}

We fitted the obtained visibilities for each spectral pixel independently with a binary model of the form
\begin{equation}
%    fit = (1.d0+p[2]^2.+(2.d0*p[2]*cos((2.d0*!DPI/wl)*(x*p[0]+y*p[1]))))/(1.d0+p[2])^2.
   \left| V_{\text{model}}(u,\upsilon)\right|^2 = \frac{1+f^2+2f\cos{\left(\frac{2\pi}{\lambda}\vec{B}\cdot \boldsymbol{\rho}\right)}}{(1+f)^2},
\end{equation}
where $\lambda$ is the wavelength of the corresponding spectral pixel, $f$ the flux ratio of the stellar components, $\vec{B}$ the baseline 
vector, and the separation vector is defined as $\boldsymbol{\rho}=\binom{\alpha_1-\alpha_2}{\delta_1-\delta_2}$ with 
$\Delta\alpha=\alpha_1-\alpha_2$ and $\Delta\delta=\delta_1-\delta_2$ as the sky coordinate offsets of the companion. It was assumed that the stellar 
surface of all components is unresolved. For the fit $\Delta\alpha$, $\Delta\delta$, and $f$ were treated as free parameters. 
Figure~\ref{fig:binmodelfit} shows the fit of the model to the measured visibilities of HD~155826. For comparison, we show the predicted visibility 
variation based on the published orbit (see Table~\ref{tbl:sci}). Differences between the prediction and actual fit are assumed to be caused by the 
fact that the astrometric measurements cover only $\approx$60\% of the full orbit (see Fig.~\ref{fig:hd155826}). The values of the individual 
spectral pixels in Fig~\ref{fig:binmodelfit} are only shown for completeness. The spectral pixel number ``1'' ($\lambda\approx2.0\,\mu$m) and ``5'' 
($\lambda\approx2.4\,\mu$m) of the four channels have a lower S/N because they receive up to 80\% less flux compared to the other spectral pixels. 
Thus, for fainter $K$ band targets, these pixels might not be useful for visibility measurements and preventing, for example, the fit of a spectral 
energy distribution over the spectral range of FSU-A. The new astrometric values for 
HD~155826 and 24~Psc are summarized in Table~\ref{tbl:astrometry}. The position of the new FSU-A measurements with respect to the published 
astrometric orbit by \citet{mas10} are shown in Figs.~\ref{fig:hd155826} and \ref{fig:24psc} for HD~155826 and 24~Psc, respectively.

\begin{table*}[!ht]
  \caption{Results of the fit of a binary model on FSU-A visibilities. The predicted values based on the orbital elements published by \citet{mas10} 
are provided for direct comparison. \label{tbl:astrometry}}
  \centering
  \begin{tabular}{lcccc}
  \hline\hline
  Parameter & \multicolumn{2}{c}{HD~155826} & \multicolumn{2}{c}{24~Psc} \\
	    & predicted & measured & predicted & measured \\
  \hline
  $\Delta \alpha$ / [mas] & 54.97$^{+3.48}_{-3.82}$  & 62.50$\pm$0.73 & -15.81$^{+2.96}_{-3.34}$ & -23.41$\pm$5.15 \\
  $\Delta \delta$ / [mas] & -51.79$^{+23.29}_{-18.71}$ & -55.30$\pm$0.58 & -60.01$^{+1.64}_{-1.36}$ & -65.50$\pm$3.78 \\
  flux ratio $f$          &  & 2.89$\pm$0.15 & & 1.25$\pm$0.61 \\
  
  \hline
  \end{tabular}
\end{table*}

%______________________________________________________________

\section{Discussion and conclusions}\label{sec:conc}
In this paper we presented the new MIDI+FSU-A observing mode with FSU-A acting as an external $K$ band fringe tracker that sends offsets to a delay 
line in real time for compensating variations in OPD caused by the turbulent atmosphere. The fringe-tracking by FSU-A is done on the science source 
itself (on-axis fringe-tracking). The $K$ band group delay and phase delay was used to predict the relative change in differential water vapor column 
densities allowing computation of the respective $N$ band values. A comparison between predicted and independently measured group delay values 
with MIDI yield residuals of less than 1\,$\mu$m or better than $\lambda/10$. With this method we were able to integrate the MIDI fringes coherently 
and to decrease the detection limit of MIDI down to 500\,mJy on the ATs and down to 50\,mJy on the UTs, respectively.\\
From our data set we estimate limiting magnitudes for the ATs of $K$=7.5\,mag for the fringe-tracking under good conditions (DIT=1\,ms, 
seeing$\leq$0.6$\arcsec$) and $N$=4.7\,mag ($F_N$=500\,mJy) for the AT case. We cannot provide limiting magnitudes for the UTs because we do not have 
a comprehensive data set as the ATs.

The scanning mode of the FSU-A was used to explore the possibility of visibility measurements in the $K$ band. To test the reliability of the FSU-A 
visibility measurements, two binary stars with previously known astrometric orbit were observed during the course of about three hours to cover large 
changes in visibility. Using a wavelet analysis, the integrated power over OPD and wavenumber were used to construct a transfer function to finally 
compute the visibilities. The precision of the measurements achieved was $\approx$2\% for the best data set. The limiting correlated magnitude for 
the 
scanning mode is similar to the fringe-tracking, i.e., K=7.5\,mag.\\
The only other instrument capable of measuring $K$ band visibilities at VLTI is AMBER \citep{pet07}. According to the AMBER user 
manual\footnote[5]{The manual is available for download from \url{http://www.eso.org/sci/facilities/paranal/instruments/amber/doc.html}} a 
typical accuracy of 5\% is achieved for visibility measurements when the $H$-band fringe tracker FINITO \citep[e.g.,][]{leb08} is used. In 
low-resolution ($R\approx35$), targets as faint as $\sim$6.5\,mag can be observed with the ATs. Given the low spectral resolution of the FSU-A and 
assumptions made to derive visibilities (see Section~\ref{sec:visobsdr}), it is certainly not an alternative for AMBER for detailed studies, but 
provides $K$ band visibilities simultaneously to $N$ band visibilities.

The MIDI+FSU-A mode offers unique data sets in the $K$ and $N$ bands with the possibility of measuring visibilities in these two spectral bands 
almost 
simultaneously. The use of FSU-A significantly increases the sensitivity of MIDI, making numerous additional sources accessible on the ATs. In 
addition, an external fringe tracker enables the observation of resolved $N$ band targets on long baselines. This and further MIDI+FSU-A data sets 
will be also a valuable source of future new-generation instruments to arrive at VLTI, such as MATISSE (Multi-AperTure mid-Infrared SpectroScopic 
Experiment, \citet{lop06}), to explore the usage of near-infrared fringe-tracking for $N$ band beam combiner.

% usage: resolved N-band targets, faint N-band targets, higher precision, simultaneous K and N band visibilities
% \citet{mue10} can use this to point out that FSU-A fluxes are higher than MIDI alone

%______________________________________________________________

\begin{acknowledgements}
The commissioning of the MIDI+FSU-A mode was a concerted effort of the European Southern Observatory and the Max Planck Institute for Astronomy, 
Heidelberg.
The research leading to these results has received funding from the European Community's Seventh Framework Program under Grant Agreement 312430. 
This research made use of the SIMBAD database, NASA's Astrophysics Data System Bibliographic Services, and of the Washington Double Star Catalog 
maintained at the U.S. Naval Observatory. Wavelet software was provided by C. Torrence and G. Compo.
%, and is available at \url{http://atoc.colorado.edu/research/wavelets/}
\end{acknowledgements}

%-------------------------------------------------------------------

\bibliographystyle{aa}
\bibliography{references}

\end{document}